# A RESTFUL APPROACH FOR MANAGING CITIZEN PROFILES USING A SEMANTIC SUPPORT

Luis Álvarez Sabucedo and Luis Anido Rifón

Telematics Engineering Department, Universidade de Vigo, Spain

{lsabucedo,lanido}@det.uvigo.es

## ABSTRACT

*Several steps are missing in the current high-speed race towards the holistic support of citizen needs in the domain of eGovernment. This paper is focused on how to provide support for the citizen profile. This profile, in a wide sense, includes personal information as well documents in possession of the citizen. This also involves the provision of those mechanisms required to publish, access and submit the convenient information to a Public Administration in due curse of a transactional services provided with the last one. Main features of the system are related to interoperability and possibilities for its inclusion in a cost-effective manner in already developed platforms. To make that possible, this approach will take full advantage of semantic technologies and the RESTful paradigm to design the entire system. The paper presents the overall system with some notes on the deployment of the solution for its further reuse in similar contexts.*

## 1. INTRODUCTION

The domain of eGovernment is a quite incipient area or research and development. Even during these last years, a large effort has been devoted to the provision of new solutions and the design of holistic frameworks; a large effort is yet to be devoted. In this road towards a fully transactional environment for services provided by Public Administrations and Agencies (PAs hereafter) there is still some gaps missing to be tackled. And, in order to properly face these challenges, it is required to keep in mind the current needs of citizens and governments such as world-wide interoperability, accessibility, and attention to security concerns.

In particular, this contribution is concern with the provision of support for the management of the profile of citizens. In the globalized context we are involved in current days, it is quite normal to get related with a number of PAs from a number of different scopes. In their regular operations with PAs, citizens are requested to provide with personal information and documents, and, in exchange, they are supplied with a number of documents. It seems clear the need to tackle this issue in a convenient manner in order to avoid unnecessary resubmission of documents and to properly manage the entire pool of documents and personal information.

In a search for a suitable approach, we must keep in mind the definition and the aim of eGovernment. As defined by UN, eGovernment is[1] "the use of information and communication technology (ICT) and its application by the government for the provision of information and basic public services to the people". EU[3] defines eGovernment as "the use of ICT in public administrations combined with organizational changes and new skills in order to improve public services and democratic processes and strengthen support to public policies". From the point of view of the World Bank[2], eGovernment can be understood as "the use by government agencies of information technologies (such as Wide Area Networks, the Internet, and mobile computing) that have the ability to transform relations with citizens, businesses, and other arms of government".

From the review of all of them, it is clear that the aim pursuit in the domain is not merely to replace current fashion systems with similar systems that take advantage of the ICT support. Thus, in the context of this contribution, it is not acceptable to go for a solution based on just allowing citizen to re-submit documents each time they are requested to do so. On the contrary, in the light of this concept for eGovernment, it is mandatory to put the citizen in the centre of processes and redesign, if necessary, all workflows to make easier things for them.

In this line, the motivation for this work comes from the observation of current fashion solutions that, in order to make the most of its potential requires the citizen to provide with personal data and a large range of personal documents. This would be the case of searchers of services, complex services that require citizen to provide data before an actual invocation is completed, or even those services that need before hand information to decide about the eligibility for a certain operation.

And, precisely, one of the burdens of dealing with Public Administration is due to the multiplicity of mechanisms to access data or to login, or the number of times required uploading the same data[6]. To fight back this issue, this paper proposes a mechanism to simplify these operations. The main aim of this contribution is to support in a holistic manner the entire profile of the citizen, including the documents he/she may in possession of. This support must be planned in such manner it would be simple to access for both, the citizen, i.e., the owner of the data, and the PA involved, i.e., the service provider.

The first step is to get a clear idea of what a document is and how it must we managed in the domain of eGovernment (see Section 2). The next step is to set the conditions and requirements for our proposal (see Section 3) and check similar initiatives on the same and related fields (see Section 4). In our case, in order to increase the interoperability of the solution, we take full advantage of semantics. This technology, which can be used for multiple purposes, will be applied to boost the interoperability of the system and to support the recovery of information. Basis of this technology are presented (check Section 5). Additionally, we are interested, not just in provide contents interoperable, but also accessible, in terms of simplicity of use for human users and third party software agents. That is the reason to deploy a simple-to-use and open software architecture. In our case, the RESTful approach (deeply discussed in Section 6) was considered the most convenient solution. An application of the semantic techniques to this particular problem is presented on Section 7. Afterwards, on Section 8, a deeply discussion about how the system works is presented.
Information regarding the design and the implementation of the final solution is presented to the reader in order to make possible its replication in further developments (see Section 9). Finally, after the work carried out, some conclusions are presented to the reader on Section 10.

## 2. DOCUMENTS IN EGOVERNMENT

The use of documents on public administrations drives the entire workflow on a Public Administration. As stated in[4] , "Government organisation need to manage electronic records as a valuable corporate information resources by constructing interacting systems of software, standards, policies, procedures and interfaces". In fact, in both eGovernment and "regular" Government, documents act as the proof for operations already performed or to be performed and compel PAs and citizen to comply with its contents.

Nowadays, the transition from paper-based administration to a digital model is on the agenda of most countries. The first step is to set the legal framework, also for this sub-area of the domain. And this new legal framework will introduce high impact changes on the day-to-day life of the citizen. Using the Spanish regulation[5], that is being published at the moment of writing this document, as a predictor for other legislations, it is possible to outline some relevant advantages for the citizens:

- Digital copies of a document will have the same value that the original ones. These copies can generate new copies.
- The use of metadata is explicitly considered on this new generation of documents. And, its content is an active part of the document itself.
- The digitalization by optic means of a pre-existent document is also considered a digital document with all the legal considerations attached.
- PAs will be in charge to preserve the documents that they issued.

These changes regarding the documents themselves will also imply the generation of a software infrastructure for its management. According to [4], a good electronic record-keeping requires, among other features:

- A clear understanding of the nature of electronic records.
- Ensure the integrity and reliability of electronic records
- A strategy to ensure that electronic records will remain accessible and usable as long as they are needed.

The considerations for developing a system to manage records or documents in the domain of eGovernment must be consistent with three different levels:

- At organizational level. All involved PAs in the system must take into account the general philosophy and comply with the guidelines established to ensure the coherence and the good performance of the system.
- At the record management level. All records generated in the system must comply with the agreed procedures and the conventions set for this porpoise.
- At the IT system level. All technical elements involved in this task must be able to work together to deliver the proper service. The must common, and almost compulsory, way to achieve this by means of open technologies and standards.

## 3. REQUIREMENTS OF THE SYSTEM

As mentioned in the introduction, currently there is a barrier for the engagement of citizens when they tackle the use of services in the domain: the time-consuming task to upload or to complete its personal profile before they access each platform they may need to interact with. In this context, when we speak about the profile of a citizen we are referring to the sum of the information regarding the citizen itself and the documents he/she actually owns.

In the given context there is a set of premises that we must consider such as interoperability of contents, access to information under secure connections, simplicity of use in the mechanisms and so on.

From the study of the requirements, it is possible to identify some use cases that will describe its functionality from the point of view of the citizen:
- Manage Documents. This use case supports the managing of all documents on the behalf of the citizen. It will include use cases to add new documents or to remove them. Also, it must be supported the functionality to recover the list of documents and one particular document the citizen may be interested at.
- Manage Personal Data. This feature allows the citizen to update and check its own personal information. Main pieces of information are name, surname, citizenship, full address, civil status, email,
- Export/import records. This feature supports the generation of a package containing all data in the system and its restoration in the same or different instance of this agent.
- Allow Access. This use case supports the access to data by a certain Public Administration once the citizen has allowed it.

Using these concepts, it is clear that the system must involve two main parts:
- On one hand, it must be provided a highly interoperable definition of the profile of the citizen, including his/her personal data and his/her documents. As show later on, this will be achieved by mean of ontologies (see Section 5 ).
- Also it must be provided a software solution that implements all features identified in this section. This implementation must comply with some compulsory features such as openness, simplicity and security. Thus, the use of a RESTful approach was decided (see Sections 6).

## 4. RELATED WORK

As previously mentioned, this contribution is not aimed to provide a final solution but to support further projects that may make good use of the possibilities of this work; that would be the case, for example, of project requesting the citizen to provide with external and additional documents. This sort of solution is not very common in independent institutions or small administrative entities due to obvious reasons of cost and actual utility as further projects are required to actually get back the investment for the project. Even there are a number of solutions that actually manage the profile the of citizen in order to fulfil services as expected, e.g., services for paying taxes; there is a lack of projects that are actually focussed on the provision of services related to the profile of the citizen and offer access to third parties.

And this is due, regretfully, to the long-term effort required to develop this sort of project and to achieve the required critic mass of users needed to make worthy the attempt.

Nevertheless, more mature domains have already tackle similar concepts. eLearning is one of those domains where researchers have devoted a large amount of resources to address this feature. In the context of this eTechnology, profile is linked to the metadata description of abilities and capacities of the learner and there are recommendations and standards for its full definition and management[7].

In our scope, anyhow, we can use some basis and concepts already deployed in the frame of other non-highly related projects. Maybe, most relevant initiatives come from the use of metadata to make annotations on documents. In this research line, we can underline some initiatives such as eGif [8] or NZeGIF[9]. These ones are oriented to provide a certain set of metadata that can be used to describe documents belonging to a certain citizen. It is also needed to outline the contributions made by CEN, mainly by mean of the CWA 14860 [10]. This document presents a metadata application profile based on Dublic Core for eGovernment for Europe. Thus, several official recommendations are considered and a mapping between different definitions of metadata is proposed for the shake of interoperability.

## 5. SEMANTICS TO DEFINE KNOWLEDGE

At this point, it is required to choose a technology, a tool, that allow us to provide some services required in this application. We would need to provide the system with support to define the information in an interoperable manner, to allow the system to discover new pieces of information from the already provided contents and to reason about the already existent information. It seems clear to us that the most suitable tool in this context is Semantics.

The semantic web has emerged as a new promising technology aimed at addressing information instead of data, i.e., it enables software agents to treat data in a meaningful manner. Making this possible would allow new mechanisms to operate on a higher level of abstraction. Also, by means of this technology, it is possible to express knowledge in a formal and interoperable way. These features will support the provision of the services claimed in this paper.

For the definition of knowledge as it is intended in the context of semantics, ontologies are a key element. In the literature, several definitions or approximations to the concept of ontology are provided. Gruber proposed a quite suitable definition for ontologies [11]:

> An ontology is a formal, explicit specification of a shared conceptualization of a domain of interest.

By means of this definition we are addressing an ontology as a support to put in a concrete way abstract information about a certain domain by means of a machine-understandable data format.

To express an ontology in a formal manner different languages are at our disposal[12]. OWL (Ontology Web Language) [13], is the W3C Recommendation intended to provide a fully functional way to express ontologies. To make different levels of complexity possible, OWL provides different sublanguages with increasing expressivity: OWL Lite, OWL DL and OWL Full. By using OWL, we are addressing a W3C recommendation that can be considered as solid and interoperable support for the provision of this solution.

Currently, there are available for developers, software libraries that support the use of semantic features required for this kind of projects. Nevertheless, we must keep in mind one of weak points of semantics: its immaturity. Even semantics appear several years ago, the software platforms, and even standards themselves, are not in a mature state and problems on software available may impact on the time-to-market of solutions.

## 6. THE RESTFUL APPROACH

As mentioned before, in order to deploy a convenient software platform, it is required to make a wise election of an architectural style, i.e., the guidelines and the principles that will rule the actual software architecture for the prototype. In the frame of this work it was decided to make use of the REST (*REpresentational State Transfer*) philosophy[14]. Systems fitting in this principle are called to be "RESTful" and, taking the client-server model as the basis, it poses some principles:

- Resources to be used always must be able to be identified from the outside of the solution. One of the aims of this sort of architectures lays on managing data, i.e., resources, regardless of their storage, origin or status. Thus, it must be provided in a unique manner to establish a link with them. The preferred way is by means of URIs (Uniform Resource Indicator) and IRIs (Internationalized Resource Identifiers). As a result, it is possible to get an unlimited number of resources.
- Messages, i.e., commands, must be a set of clear and simple verbs. Contrary to what happen to resources, actions must be limited and well known. Usually, these systems are supposed to be based on CRUD (Create, Read, Update and Delete) systems. Invoking these commands on resource within the system, all required operations should be possible.

On the light of these ideas, it is clear that HTTP is the perfect example for RESTful system. It is possible to use a limited set of commands (addressed in the HTTP1.0[15] or HTTP1.1[16] specification) on any URL provided by the software agent in use. Anyhow, RESTful systems do not have to be HTTP or even based on HTTP.

The purpose for a RESTful system is to manage all resources in the system by means of this set of limited and well-know *verbs*. Reader can compare this feature with RPC-based system where the set of operation is freely upgradable and parameters attached do not identify the resource but the properties to be changed.

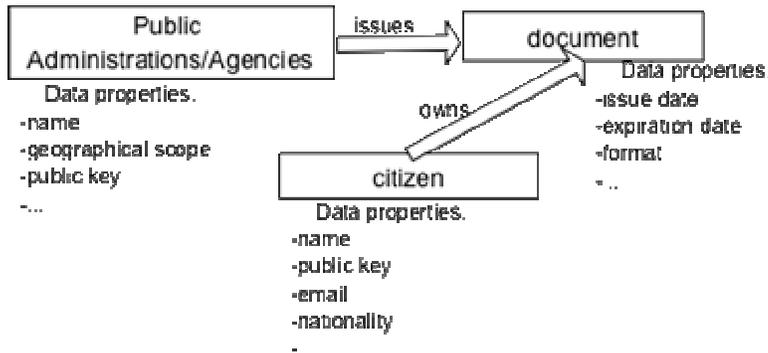

*Figure 1. Excerpt of the ontological model*

The change of the status for resources in the RESTful approach is driven by these verbs for actions and never stored on the server that will not have memory about former invocations. This feature drives the design of the servers and makes possible a higher scalability and larger simplicity for these solutions.

In our case, as mentioned before, we will take full advantage of semantics. Thus, instead of providing links to "simple" resources, semantically defined items will be used. Thus, the actual content to be managed by the system belongs to an ontology. This feature enables the system to provide with additional features not present in non-semantic environments such as reasoning about the outcome of an operation, fully autonomous discovery of next iterations, or smart searches for additional services.

## 7. SETTING THE CONTEXT

Taking advantage of semantics is not a trivial task and it requires a large effort. The very first step is the provision of a ontology that can put us in position to express in a explicit manner all the relevant knowledge to characterize the environment of the proposed solution. This task requires a solid methodology that provides us with a systematic and well-documented way to achieve this goal. Several methodologies are possible: Cyc[17], Kactus[18], On-To-Knowledge[19], etc.. But our decision was to make use of Methontology0. Main reasons for this decision are due to its background for large projects, support from FIPA[21] and previous expertise.

The very first step is to make clear which are the goals of the provided knowledge support in the system.

To make that explicit the competence questions is the chosen tool. In our case we would need to be able to answer, among others, the following questions:

- What documents belong to a citizen?
- How can a certain document be recovered?
- Where does the citizen currently live?
- How can the citizen be addressed?
- Is a certain document still valid?

Taking these questions as a starting point and following the steps described by Methontology, an entire ontological system is generated. In due course of this methodology, a number of phases are undertaken: specification, conceptualization, formalization, and implementation. The last phase identified in the methodology, maintenance, is to be performed along the entire life of the system.

The final result is the complete ontological model of the system. This holistic model of the domain is made of different modules provided under the form of an independent ontology. Therefore, different agents for different porpoises can just pick the piece of the ontology related to its goals. Additional, this allows us to reuse pieces of ontology already developed and commonly accepted such as FOAF[22] or the W3C Time ontology[23].

As an excerpt and for the sake of simplicity, the most relevant part of this ontology for our porpoises is presented on Figure 1. In the figure, it is shown how the different classes are related. Besides, the class document includes several sub-classes for national, international, regional and personal documents. Each document included in the system will be translated as a new instance belonging to the appropriated class.
Main classes identified within this proposal are:

- Citizens. This class models the information regarding citizens, the actual owners of documents in the system. As mentioned previously, FOAF model is used on the characterization of this class.

- PA. Public Administrations, as the service providers in this schema, play a main role in this proposal and a complete model of their abilities and capacities must be included.

- Document. This is the central element of the ontology and a complete description of this item must be provided. Actually, this class involves information regarding a different number of issues in order to enable to locate the required document using the maximum number of parameters.

These classes defined within the ontology are related through several properties:
- issues. It links a document with the PA that issued it.
- owns. It links a document with its owner.
- expires. It states the date of expiration for a document.
- and many others that enables the system to perform inferences and discovery of new information.

## 8. SETTING THE INTERACTION MODEL

| HTTP command | Resource | Meaning |
|---|---|---|
| GET | Document | Recover a particular document within the system |
| POST | Citizen | Update all the personal data about a citizen |
| PUT | Document | Add a document to the system |
| DELETE | Document | Remove a document in the system |

*Table 1.- Verbs and its meaning in the proposal*

As mentioned in previous sections, the use of a RESTful approach poses some restrains about how operations can be deployed and which kind of information can be used. From Section 7, a characterization of resources, i.e., documents, PAs, and citizens, is provided. Besides, these resources, as they are defined in terms of an ontological support, can be addressed by means of a link. This is due to the fact that all instances in an ontology are provided with an identifier under the form of a URL.

In order to fully characterize the architecture itself of the system, it must be identified how interactions are conducted and which operations are possible. These operations will be mapped into *verbs*, using a RESTful terminology, and used with the corresponding resources. In the case of this proposal, we will stick with the HTTP protocol and take advantage of the provided commands. These operations will be used with a concrete meaning. On Table 1 some combinations of *verbs* and *resources* are displayed. The reader should note that the same verb or command applied to different resources will have a different meaning. Upon each invocation of the former commands, the server will respond with information expressed in the terms of the ontology included in the system, i.e., a rdf-like file will be generated and sent back to the client. Thus, the agent receiving this response will be able to understand the knowledge implicit in this response and, if the case, take the correspondent decision.

As all the operations possible are defined in terms of the provided ontology it is not required to negotiate about the interface of the service or decide and discover how a particular WS must be addressed.

The server in this approach will be administrated by a reliable entity, such as a Public Administration or lawyer-like agent provided with the required legal support. The software to fulfil the duties of the client can, in theory, be provided by third parties as far as they stick to the defined interfaces. As already mentioned, it makes no sense to keep this sort of service hidden for citizen or prevent them to know about how their data is actually managed.

Taking into account these main guidelines, functionalities identified in Section 3 must be faced. For each one of them, a number of considerations must be done.

### 8.1. Manage Personal Data

In the provided system, the server entity stores on its internal DBMS all the information regarding the citizen. This information is compliant with the schema defined in the ontology previously addressed in this paper. From a practical point of view, the user must submit a HTTP request using the method POST on a resource "citizen". This resource it is an entity of the class that the server will accept and update the information conveniently on its internal database.

### 8.2. Manage Documents

This is a feature relevant in the system, as it must deal with documents on the behalf of the citizen. This involves tasks related to two different features. The system takes care of the digital version of document regardless of its origin and stores it in the server where the system is currently running. Besides, it manages the knowledge associated to each document. The user must provide this information, as the system is not capable of inferring that information from the file itself. Once all these pieces of information, including the file itself and the related data expressed in terms of the ontology, are provided, the system takes care of them and stores them. Regarding to recovery issue, all this related information expressed with the ontological support is at the disposal of the citizen to find a particular document he/she may be interested at.

Besides of the already mentioned functionalities for managing documents, the system can also provide with enhanced characteristics regarding the use of the information within it. Some advanced features obtained from the application of semantic tools are:
- Discover those documents that are not longer valid. And also, discover in advance when some certain document will expire and notify the citizen. This would be of special interested for documents, p.e., such as the passport for ahead planning travels. To implement this feature, SPARQL queries that will look for documents that fit a particular pattern are used.
- Find out duplicated documents from different PAs. Using the properties defined the system identifies those documents that could be duplicated or redundant and let the user know about them.
- Change the status of the citizen regarding certain features such as the civil status. Using the properties of the documents and SWRL rules we can reason about documents and find out new pieces of information.

### 8.3. Export/Import data

The citizen has the possibility to pack and unpack all the information that comprises his/her personal record. In our case, this will involve the ontology containing his/her personal data and also the files where their documents are stored. To make that possible, the prototype generates a zip file containing the instance of the ontology needed to define entirely the citizen profile and also the documents already uploaded. The reverse operation is also at the disposal of the citizen that, making use of a web form, will be in position to upload a zip file and insert that content on the system as a new instance of his/her records. It is important to note that at this step, the ontology provides a common language to define the contents to be used; therefore, the interoperability of the so-generated file is guaranteed.

### 8.4. Allow Access

This use case solves a scenario where a citizen wants to put a particular document at the disposal of a certain Public Administration. In this case, the citizen chooses which document a particular Public Administration can access. And, for that single document, it is generated one single URL that drives the access to the previously selected document. In order to guarantee the privacy of the citizen, this URL will be accessible just under certain conditions:

- The link will be operative just once. Afterwards, the document will not available at that URL.
- The link will be accessible only from a certain range of IP addressed previously attached to the particular PA allowed to get the data.
- The link is operative a certain amount of time. Afterwards, regardless of the use done, it will be not longer accessible.
- The link will contain at least 24 random characters to prevent from inappropriate access attempts.

This functionality shows the balance between the power of the software provided and its integration with external platforms. It would be a better approach for the shake of interoperability to directly allow external software to access documents within the system. Nevertheless that approach would require a larger effort for PAs willing to take part in the proposal. Under this scheme, with no cost for external applications, it is possible for them to take part in the system as they are just expected to access a regular URL where the desired document is hosted.

## 9. IMPLEMENTING A PROTOTYPE

In order to test the solution proposed, a prototype was developed. This prototype is composed of two separated software agents. On one hand it is provided the server of the system responsible to fulfil the request of service performed by the citizen and also to conveniently manage the data already uploaded to its pool of contents. On the other hand it is required to provide an implementation of the client capable of invoking the services previously identified. This agent software will act mainly as a front end for the functionalities provided by the server as nearly neither data is stored on it, nor responsibility is put on it.

The implementation of this system must ensure the security of the system from different points of view. To undertake this feature, compulsory due to the nature of the domain, several measures were planned:
- Records of documents include a hash field to ensure that information has not been anyhow altered. Thus, agents involved must check the conformance with this code previously to each sensitive operation.
- Citizen must prove their identity using a PKI infrastructure[25]. This is the method to ensure the authenticity of the user and also the manner to avoid the repudiation of the operation. Depending on the country and the available infrastructure, this certification will be provided in a different manner.
- The communication among agents in the system will be conducted always using secure channels using the SSL protocol.

For the shake of simplicity, all these features will be taken for granted and further description of the prototype will be built on the basis of these premises.

This software agent must make good use of the described semantic support and, of course, meet the actual goals of the system as they were planned originally. To this end, Java[24] was the chosen option as the software platform. Due to its support for large-scale projects, its community of developers, and support for different and required libraries, this select technology was considered the best option to develop the software prototype.

To manage the semantic contents within the server, Jena[27] was chosen as the library to deal with the involved knowledge. As it was shown previously, we would require to undertake a number of operations on the semantic information within the system. These operations involve the execution of SPARQL queries, adding/removing knowledge to the system, etc. Jena provides support for those operations in a simple manner. In order to illustrate the behaviour of this software tools it is included the snippet required to find out which documents belong to a certain citizen using the suggested tools:

```
// The first step is to load the model of the ontology
   OntologicalModel individuals = ModelFactory
      .createOntologyModel(PelletReasonerFactory.THE_SPEC);

   // Afterwards, the actual contents must be read
   individuals.getDocumentManager()
   individuals.read("http://portfolio.det.uvigo.es/ontology.owl");

   // The query for the desired knowledge must be prepared
// In this case, the query will request documents belonging to a certain citizen
   String queryString=Portfolio.generateQuery();
  Query query = QueryFactory.create(queryString);
```

```
     //The actual query is introduced
  QueryExecution qe = QueryExecutionFactory.create(query, individuals);
    //An object to launch the query is created
  ResultSet rs = ResultSetFactory.copyResults(qe.execSelect());
    //And finally, the query is launched and the results copied to another container
  qe.close();
```

Reader should take into account that for each document uploaded to the system, it is required to perform two different operations:

- Firstly, it is required to add a new piece of ontological knowledge
- Later, the document itself must be uploaded to the server to keep a copy of it for its further use.

The latter operation is quite simple, as it is possible to upload it in a simple manner using a HTTP channel. Nevertheless, to perform the second operation it is required to manage the ontology in the system.

```
// Defining namespaces of concepts to be used
   String RDF="http://www.w3.org/1999/02/22-rdf-syntax-ns#";
   String Citizen="http://portfolio.det.uvigo.es/citizen.owl#";
   String Document="http://portfolio.det.uvigo.es/document.owl#";

   // Generating an ontology model and loading data
   OntModel individuals = ModelFactory.createOntologyModel();
   individuals.read("http://portfolio.det.uvigo.es/data.owl");

   // Generating an ID for the new LE
   String ID = Document  + PA.toString() + registryInternal.getNext().toString();

   // Creating of the new instance for the Document to be inserted in the ontology
   Resource doc1=individuals.createResource(ID);

  // Defining the properties for doc1
                  doc1.addProperty(individuals.getProperty(Document+"Name"),
      individuals.createTypedLiteral(name,
                        XSDDatatype.XSDstring));
           //Further pieces of information can be added in a similar manner

  // Finally,  new contents are put on the ontology in a permanent manner
  try {
  FileOutputStream fos = new FileOutputStream("/dataPool/data.owl");
  individuals.write(fos, "RDF/XML-ABBREV",
       "http://portfolio.det.uvigo.es/data.owl");
  fos.close();
  } catch(IOException e) {
  e.getMessage();
  }
```

Dealing with the communication among agents in the system is quite simple as we realy on the basis of the RESTful paradigm. Therefore, it is only required to open an HTTP connection and compose the correspondent request using the proposed matching of commands and resources.

The implementation of the client agent tackles mainly the provision of a front-end for the invocation of services provided by the server. This client agent is actually responsible only for the presentation of functionalities but nearly no sensitive or resource-demanding operations are required to meet its requirements. For this proof-of-concept implementation, it was decided to develop an entire application but further implementations of the system could be presented under the form of a simple web site including all services required.

This client was developed under the philosophy of a light-weighted client. Therefore, nearly no logic operation is performed on the client-side. And, also, for the shake of simplicity, semantic operations are also conducted on the sever side. Thus, the client just sends the pieces of information and the server is responsible for constructing semantic queries and analyzing the requested operation.

## 10. CONCLUSIONS

eGovernment is getting more mature as more resources are devoted to the domain and more experienced become developers. These new and more complex projects require not just support for final solutions but also supporting projects that may enable the access to these final services. In the frame of these projects, an efficient and simple manner to handle the profile of the citizen is required. And that is the motivation of the proposed system, to overcome some of the burdens attached to projects in both, conventional government and eGovernment:

- Different login mechanisms or passwords to access documents for the same user. As we provide with one single entry point for accessing the citizen profile, he/she does not need to use different login systems by means of different smart cards or other authentication mechanisms.
- Holistic managing of citizens' data. Classifying, searching, and storing personal documents is a time-demanding operation simplified by this tool. Citizens may get an overwhelming amount of documents as results of different operations in public administrations. This may turn out into a burden for them that can be avoided making use of the proposed implementation.
- Securing the access to information. Involved documents in these operations usually hold personal data. As they are dealt with in the proposed architecture that is provided with reliable security measures, no concerns should arise regarding this point.

This sort of support is intended to be used from different projects in order to provide synergies that make this solution actually appealing for both, Public Administrations and citizen. Public Administration will get a simplified mechanism to access secure data. Conversely, citizen will get all their data securely handled, by the entity of his/her election, and ready to be used for any Public Administration. This solution also does not require any special format for documents neither imposes any particular scheme to deal with the documents itself. The Public Administration that will use the document will have full access to the documents, once the citizen agrees on that, to manage it at its own convenience.

The reader should also note that the level of integration of this solution in the internal behaviour of the PA is quite limited. Actually, PAs taking part in this proposal will have access just to raw files, besides all the semantic annotations added, in similar manner to those PAs that just consider an approach closer to the traditional paper-based methodology. And this feature, to certain extension, is on porpoise as we can not expect PAs to directly embrace directly digital procedures. On the contrary, it must be provided an acceptable path from a traditional paper-based administration to more evolved scenario with acceptable costs and procedures in the transition. The presented approach seems to the authors as the best alternative under such conditionings.

From a technological point of view, the use of semantics turned out to be a quite useful support. It provides us with a highly interoperable manner to define knowledge in the system that is actually shared among different entities with quite little linking. Also, it allows us to provide a way to recover information from different sources and to make discoveries about the current status of the citizen and to add new knowledge to the system.

The use of simple software architectures, such as RESTful, proved to meet our requisites and it would require little effort for external platforms to get involved in the solution.

The presented implementation of the platform was intended only as a proof-of-concept for the viability of the proposal. Nevertheless, some problems arose and they should be taken into account for further implementations.

A quite relevant issue involves the non-mature stage of semantic libraries. Even a large community of software developers are currently working in the area, there is no a mature and complete solutions. Besides, the management of such a large amount of semantic information is not currently a simple matter and the times for loading contents on the main memory of the system may grow up too fast for any interactive solution. Also, it must be mentioned that this prototype was build on the top of a OWL ontology; nevertheless, during the development of this proposal, W3C has released the evolution of this recommendation and currently the official recommendation from W3C is to use OWL2[30].

The actual adoption of the proposed schema by Public Administrations should involve a review of the local/national legislation in order to make sure its compliance with their particular regulation.

## ACKNOWLEDGEMENTS

This work has been funded by the Ministerio de Educación y Ciencia through the project "Servicios adaptativos para e-learning basados en estándares" (TIN2007-68125-C02-02) and by the Xunta de Galicia, Consellería de Innovación e Industria "SEGREL: Semántica para un eGov Reutilizable en Entornos Locais" (08SIN006322PR).

**Authors**

Dr. Luis M. Álvarez Sabucedo has a Telecommunication Engineering degree with honours (2001) and a Tele- communication Engineering PhD with honours (2008) by the University of Vigo. Currently, he is an Assistant Professor in the Telematics Engineering Department of the University of Vigo. His research interests include seman- tic technologies, web-based solutions and eGovernment. His results have been published in several fora such as international conferences and journals. Also, he is involved in a number of national and international research projects.

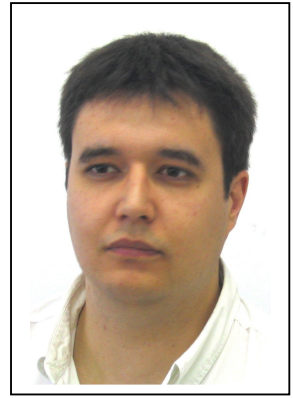

Dr. Luis Anido Rifón has a Telecommunication Engineering degree with honours (1997) in the Telematics and Communication branches and a Telecommunication Engineering PhD with honours (2001) by the University of Vigo. Currently, he is Associate Professor in the Telematics Engineering Department of the University of Vigo and holds the post of Director of the Innovation in Education Unit of the University of Vigo. He has received several awards by the W3C, the Royal Academy of Sciences and the Of cial Spanish Telecommunication Association. He has authored more than 180 papers in journals and conferences. He is also the manager of the ICT Galician Research Program, technical secretariat of the CTN71 SC36 group of the Spanish Association for Standardisation and Certi cation (AENOR) and head of the Spanish delegation to ISO/IEC JTC1 SC36.

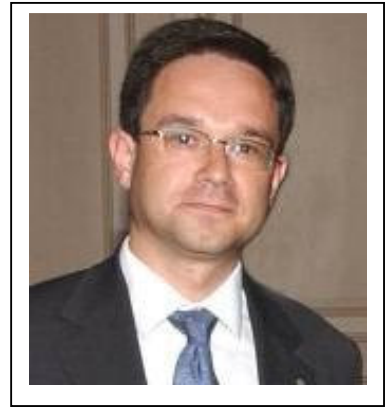